# Klinkenberg Slippage Effect in the Permeability Computations of Shale Gas by the Pore-scale Simulations

**Jun Li [1], Abdullah Sultan [1,2]**
[1] *Center for Integrative Petroleum Research,*
[2] *Department of Petroleum Engineering,*
*College of Petroleum Engineering and Geosciences, King Fahd University of Petroleum & Minerals, Saudi Arabia*

**Abstract**
The prediction of permeability (i.e., apparent permeability) for the shale gas is challenging due to the Klinkenberg slippage effect that depends on the pore size and gas pressure. A recent Monte Carlo molecular simulation method (i.e., DSBGK method) is employed to accurately compute the permeability by the pore-scale simulations at different pressures. The computed results of a benchmark problem proposed here are used to verify the accuracy of the simple Klinkenberg correlation model, which relates the permeability to the intrinsic permeability (i.e., liquid permeability) and pressure. The verification shows that the Klinkenberg correlation model as a fitting formula is appropriate for the industry applications since the relative error is small in the whole range of the flow regime as long as it has been calibrated for each particular rock sample at the two ends with low and high pressures, respectively, by determining the model parameters using accurate permeability data that can be obtained by the scheme presented herein.

## Introduction

Productive gas-shale systems consist of four types of porous media: inorganic matrix, organic matrix, natural fractures and hydraulically induced fractures. Numerous organic flakes are sparsely scattered inside the inorganic matter and intersected by natural fractures [1]. The Knudsen ($Kn$) number is defined here as the ratio of the molecular mean free path to the representative pore size, where the molecular mean free path increases with the decrease of pressure during the gas production process. The $Kn$ number is usually large in the organic matter due to the nano-pores, which makes the Klinkenberg slippage effect [2] remarkable and increases the permeability.

The traditional experimental technique to measure the flow rate at steady state is not applicable for the permeability measurement of shale gas because it requires a considerable time and the low flow speed is usually dominated by the noise in the measurements. In the current experimental study, the transient pressure decay is observed and applied in a mathematical model to inversely predict the permeability using a correction term at large $Kn$, which is obtained in a unidirectional channel flow problem and determined using the solution of the linearized Boltzmann equation or the direct simulation Monte Carlo (DSMC) method [3]. The validity and performance of the current experimental scheme are limited since the passage section of real pore space is much more complicated than that of a channel used to obtain the correction term.

Gas flow problems at different scales need different theoretical descriptions. If the pore size is comparable to the molecular mean free path but much larger than the molecular size of the fluid, the flow phenomenon and the boundary condition can be described by statistical models. Deterministic models, like molecular dynamics (MD) simulation, become necessary only if the pore size is comparable to the molecular size of the fluid. In the case of shale gas, the pores with a size comparable to a methane molecule are usually blocked due to adsorption and so negligible in the computation of permeability. Thus, the statistical models are valid to simulate the pore-scale flows of shale gas. The gas flows can be divided into different regimes according to *Kn* [4, 5]: 1) continuum regime with *Kn*<0.01 (*note*: not 0.001 as shown in the following results), where the Navier-Stokes (N-S) equation and the no-slip boundary condition are valid; 2) slip regime with 0.01<*Kn*<0.1, where a slip boundary condition should be used with the N-S equation; 3) transitional regime 0.1<*Kn*<10, where the Boltzmann equation at the molecular level in a statistical way is necessary; 4) free molecular flow regime 10<*Kn*, where the Boltzmann equation is significantly simplified and analytical solutions are sometimes available. The operating range of productive shale gas usually covers the slip and transitional regimes [5]. Compared to the pore-scale gas flows in the conventional reservoir with large pore size and thus small *Kn*, the adsorption and slippage effects at the pore surface are remarkable in the shale gas flows due to small pore size. The adsorption effect is neglected here and the current study focuses on the permeability variation with pressure due to the Klinkenberg slippage effect. The adsorption effect can be reflected by modifying the pore space according to the adsorption thickness at the pressure concerned [6], which will be investigated in future work.

Many empirical correlation models are proposed to estimate the permeability variation with pressure for shale gas as discussed in [7]. Klinkenberg first proposed a simple and first-order model, which correlates the permeability to the intrinsic permeability and pressure based on the analysis of the experimental data [2]. A second-order correlation is proposed [8] and the coefficient can be correspondingly determined [9]. Additionally, the dusty gas model (DGM) can be used to derive a different estimation model, which correlates the permeability to the intrinsic permeability, effective diffusivity, viscosity and pressure [6, 10], and the coefficients are determined for real rock samples in [7]. These correlation models for the estimation of permeability are proposed to avoid the pore-scale simulations of the gas flows at high *Kn*, which require sophisticated computational methods based on the complicated Boltzmann equation.

The permeability variation with pressure is numerically studied here by the pore-scale simulations of gas flows using a recent Monte Carlo molecular simulation method, namely the DSBGK method [11-13] based on the Bhatnager-Gross-Krook (BGK) equation that is a good approximation to the Boltzmann equation in problems with small perturbations and more accurate at high-pressure conditions than the Boltzmann equation. The DSBGK method is verified against the standard DSMC method [14, 15] over a wide range of *Kn* number in several benchmark problems but much more efficient than the DSMC method. To make the results reproducible and available for the verifications of other numerical methods, a two-dimensional benchmark problem is proposed here to study the Klinkenberg slippage phenomenon. The computed results for real three-dimensional digital rock samples with 100-cubed voxels are available and will be published separately. To the best of our knowledge, the present work is the first study that investigates the Klinkenberg slippage effect using accurate pore-scale simulations rather than correlation models.

For the pore-scale simulations, the traditional DSMC method is a standard method to obtain solutions of the Boltzmann equation at arbitrary *Kn* but its computational cost is prohibitive in the case of low-speed as occurred in the shale gas because of statistical noise [4]. The ordinary lattice Boltzmann method (LBM) applies a very rough discretization in the molecular velocity space to solve the BGK equation and its accuracy at large *Kn* can be improved by increasing the number of velocity points, which unfortunately will increase the computational cost significantly [16]. A simple but artificial scheme is to modify the effective viscosity with the spatial location by adjusting the relaxation time but the verifications are usually limited to unidirectional flows in a moderate range of *Kn* [17-19].

**Simulation Method**

The pore-scale gas flows at different pressure conditions are simulated here by the Fortran MPI software package *NanoGasSim* developed using the DSBGK method that is detailed in [12]. At the initial state, the computational domain is uniformly divided into many cells, which are either void or solid, and about twenty simulated molecules are randomly distributed inside each void cell (i.e., pore space) and assigned with the initial positions, velocities and other molecular variables according to a specified initial probability distribution. Each simulated molecule moves uniformly and in a straight line before randomly reflecting at the surfaces of solid cells according to the diffuse reflection model that is accurate for usual surfaces without polishing, including both organic and inorganic surfaces inside the shale rocks. During each time step, the trajectory of each particular molecule may be divided into several segments by the cell's interfaces and its molecular variables are updated along each segment in sequence at the moving direction. Simulated molecules are removed when moving across the open boundaries during each time step and then new simulated molecules are generated after each time step at the open boundaries according to the specified pressures. At the end of each time step, the number density, flow velocity and temperature at each void cell are updated using the increments of molecular variables along these segments located inside the concerned cell according to the conservation laws of mass, momentum and energy of the intermolecular collision processes. For each simulation at a given pressure, the volumetric velocity component along the driving direction at steady state is used to compute the permeability.

**Results and Discussions**

A two-dimensional benchmark problem is proposed in Fig. 1, where the definition of the representative pore size *R* is clear. The domain sizes at both *x* and *y* directions are the same and equal to *L*=5*R*. The computational domain is divided into many cells and we use Δ*L* to denote the cell size if Δ*x*=Δ*y* while sometimes Δ*x*≠Δ*y* is applied to investigate the influences of spatial resolution at different directions to the simulation accuracy. The time step Δ*t* is selected to make the average molecular displacement during each Δ*t* is smaller than the molecular mean free path $\lambda_0$. The pressure difference between the inlet and outlet is only 1% here such that the dependence of permeability on the pressure can be accurately studied. For each simulation at a given pressure $p_0$, the volumetric velocity component $\bar{u}$ along the driving direction at steady state is used to compute the permeability $\kappa$ as follows:

$$\kappa = \frac{\mu}{(p_0 - 0.99 p_0)/L} \bar{u} = \frac{\mu}{(p_0 - 0.99 p_0)/L} \frac{\sum_{j \in \text{void}} n_j \Delta V_j u_j}{n_0 V_{\text{all}}}, \qquad (1)$$

where $\mu$ is the dynamic viscosity, *L* is the domain size along the driving direction, $n_j$, $\Delta V_j$ and $u_j$ are the number density, volume and flow velocity component at the void cell *j*, respectively, $V_{\text{all}}$ is the total volume of the computational domain, $n_0 = p_0/(k_B T)$ is the initial number density determined using the initial pressure $p_0$, Boltzmann constant $k_B \approx 1.38\times10^{-23}$ J/K and temperature $T$ that is fixed to 300 K in all simulations. In the current study, the flow is always driven at the $x$ direction and only the permeability along the $x$ direction is computed to study $\kappa = \kappa(p_0)$. The Knudsen number $Kn=\lambda_0/R$ is used in Fig. 3 to analyze the variation feature of permeability with the pressure and $\lambda_0$ is computed as follows [4]:

$$\lambda_0 = \frac{16\mu}{5p_0}\sqrt{\frac{k_B T}{2\pi m}}, \qquad (2)$$

where we set molecular mass $m=26.63\times10^{-27}$ kg for methane and $\mu=1.024\times10^{-5}$ Pa·s for simplicity but the extension to pressure-dependent viscosity is straightforward.

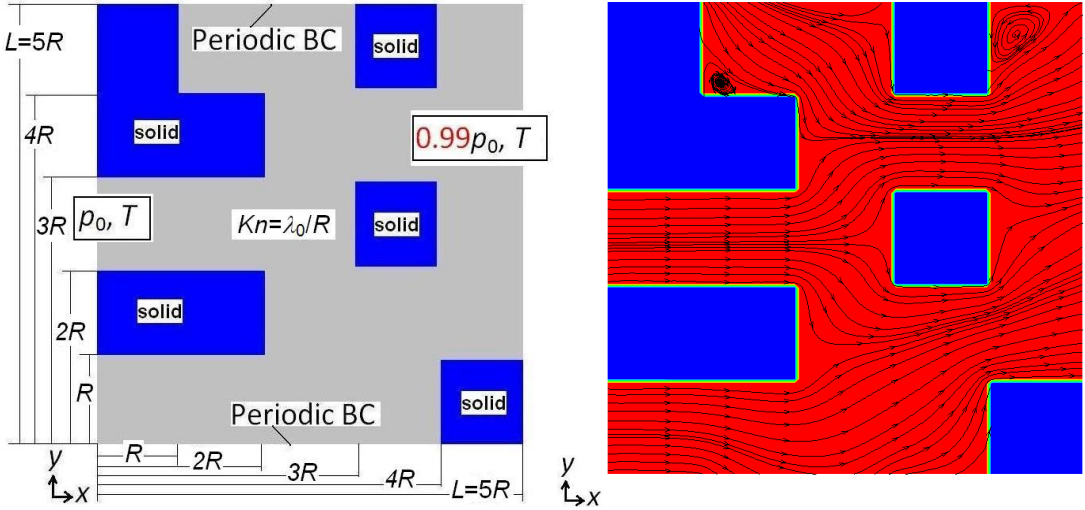

Fig. 1. Benchmark model of two-dimensional digital rock with $R$=200 nm (left) and the steady state streamlines computed using 100×100 cells at $p_0$=0.4 MPa (right) with $\lambda_0$=12.89 nm and $Kn$=0.0644.

Table 1 shows the permeabilities computed using different cell numbers for the same case. The accuracy is acceptable because the relative error is about 10% even when 10×10 cells are used for the case of $p_0$=0.04 MPa while at least 100×100 cells are required for the case of $p_0$=0.4 MPa as shown in Table 2. The general rule for selecting the cell number is to make the cell size $\Delta L$ slightly smaller than $\lambda_0$ since the change of permeability due to the computational error associated with the spatial resolution $\Delta L$ becomes negligible when $\Delta L/\lambda_0$ is less than 1 as shown in Fig. 2. Table 2 also shows that it deserves priority to increase the cell number along the directions perpendicular to the driving direction if high spatial resolutions at all directions are not affordable because of the prohibitive computational cost. Additionally, the resolution difference between different directions should be less than one order of magnitude.

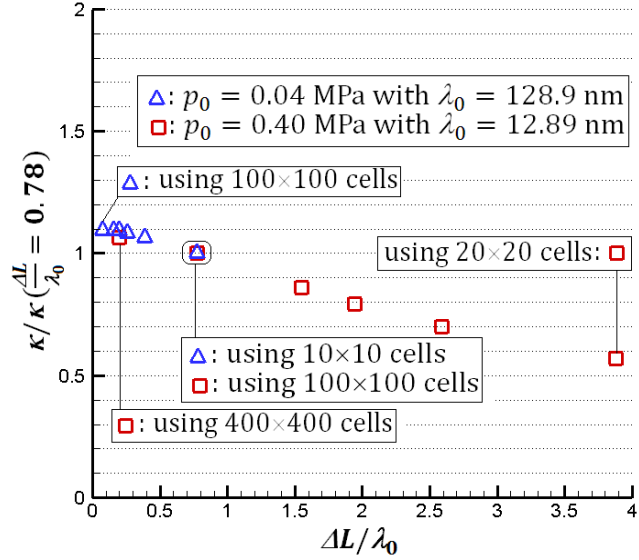

Fig. 2. Computational error due to using large cell size $\Delta L$ relative to $\lambda_0$.

Using more cells improves the accuracy but meanwhile also increases the computational cost since the average number of simulated molecules per cell is usually fixed. Based on the above error analysis, the pore-scale gas flows at different pressures ranging from 0.001 to 3 MPa are studied using different cell numbers to ensure that $\Delta L/\lambda_0$ is always less than 1. The permeability variation with the reciprocal pressure is plotted in Fig. 3. As we can see, the permeability is a constant (i.e., the intrinsic permeability $\kappa_\infty$) when the pressure is very large making $Kn<0.01$, which is consistent with the observation of conventional reservoir, where the permeability is a rock property and independent of the flow conditions and the nature of fluid. In the case of $Kn<0.01$, the numerical methods based on the N-S equation and no-slip boundary condition are valid and will predict a constant permeability as expected. At another limit with a very low pressure making $Kn>10$ (*note*: could occur earlier with $Kn>2$ as shown in Table 3 for this particular case), the permeability becomes a linear function of the reciprocal pressure, which is consistent with the constant volumetric velocity component (i.e., $\bar{u}\cong 0.36$ m/s) observed in the simulations at large $Kn$ since the pressure difference between the inlet and outlet linearly increases with $p_0$ in our simulations. This constant volumetric velocity can be understood using the kinetic theory behind the Boltzmann equation as follows: if the gas group at $0.5p_0$ obtains a volumetric velocity $\bar{u}$ when being driven with $\Delta p=0.005p_0$, the second gas group at $0.5p_0$ will also get the same volumetric velocity $\bar{u}$ when being driven with $\Delta p=0.005p_0$ and flowing together with the first group because the intermolecular collision is negligible at large $Kn$ (*note*: $Kn=\lambda_0/R>1$ implies the frequency of intermolecular collisions is lower than that of the molecular collisions with wall), consequently, the collective gas group (i.e., the combination of the first and second groups) at $p_0$ has a constant volumetric velocity $\bar{u}$ when being driven with $\Delta p=0.01p_0$. The linear relationship between the permeability and the reciprocal pressure at low pressure is always true even if the adsorption effect is considered because the adsorption thickness becomes negligible and unchanged in the limit of $p_0\rightarrow 0$. The simple Klinkenberg correlation model is proposed to predict the variation of permeability with the pressure [2]:

$$\kappa = \kappa_\infty(1 + \frac{b}{p_0}), \qquad (3)$$

which satisfies the general features at the two limits discussed above. Note that we take the Klinkenberg model as a fitting formula instead of a physical model and determine the model parameters by calibration using the computed results. This significantly improves the accuracy of the Klinkenberg model compared to the observed performances in most of the recent publications, where the Klinkenberg model was applied as a physical model and the model parameter $b$ was emperically determined without using accurate permeability data at high $Kn$. The computed permeability at $p_0=3$ MPa with $Kn=0.0086<0.01$ is used as $\kappa_\infty$ and thus $\kappa_\infty=1.993\times10^{-15}$ m$^2$. The computed permeability at $p_0=0.001$ MPa with $Kn=25.78>10$ is used to determine the parameter $b$ and we get $b\approx18.5\times10^4$ Pa. Fig. 3 shows that the permeability estimation by the Klinkenberg correlation model agrees well with the accurately computed permeability in the whole flow regime (*Note*: the relative error is within 10%). For real rock samples with irregular pore space and adsorption effect, the Klinkenberg model should also work well as long as the two parameters $\kappa_\infty$ and $b$ are accurately determined for each particular rock sample. As shown in [2] and detailed in Appendix here, the Klinkenberg model using $\kappa_\infty$ and $b$ given in Table 8 of [2] agree very well with the experimental permeability data of two rock samples given in Tables 5 and 6 of [2]. The intrinsic permeability $\kappa_\infty$ can be computed by the traditional finite-volume method (FVM) or the ordinary LBM simulation with the consideration of adsorption effect. Then, the determination of parameter $b$ depends on $\kappa_\infty$ and the permeability at a low pressure and the corresponding DSBGK simulation requires low computational cost since the refinement of cell is not necessary at low pressure with large mean free path. For industry applications, the permeability of rock sample is measured to establish geologic correlation and for quantitative calculation of the production rate. In these calculations several approximations have to be made, so that it is not necessary to know the permeability with a high degree of accuracy [2]. Thus, the Klinkenberg correlation model between the permeability and the pressure is appropriate for industry applications. Additionally, some other correlation models [3, 8] estimate the permeability using $Kn$ rather than the pressure but the definition of $Kn$ cannot be accurate in the real applications since the pore size always has remarkable variation with the spatial location. In contrast, the application of the Klinkenberg model can avoid the potential error due to uncertainty associated with the definition of $Kn$.

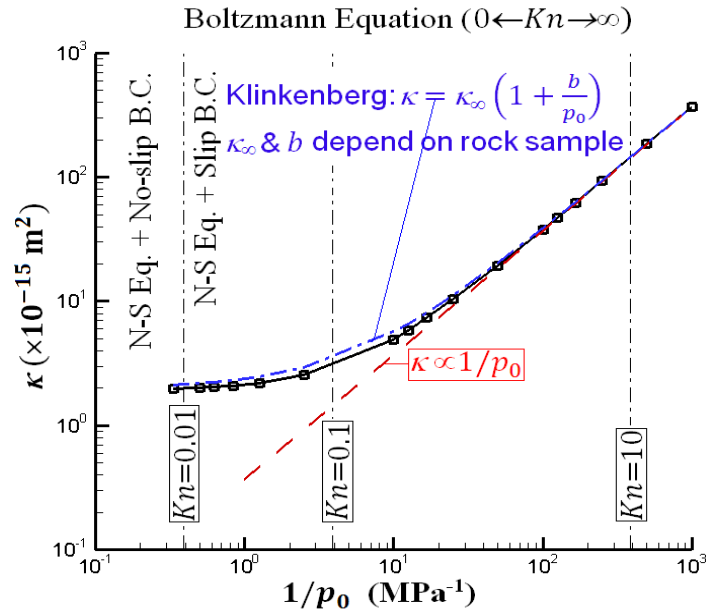

Fig. 3. Permeability variation with the reciprocal pressure; solid line: computed results by pore-scale simulations, dash-dot line: estimation by Klinkenberg correlation model.

## Conclusions

The Klinkenberg slippage effect is studied by the pore-scale molecular simulations using the DSBGK method and the permeability variation with pressure due to slippage is accurately obtained in a two-dimensional benchmark problem. The obtained results can be used as a reference to calibrate empirical correlation models and to validate other pore-scale simulation methods (e.g., LBM) based on artificial corrections at high $Kn$.

To improve the accuracy of Klinkenberg correlation model, we use it as a fitting formula instead of a physical model and the accurate permeability data obtained by our pore-scale simulations is used to determine the model parameters by calibrations at the two ends with low and high pressures, respectively. The agreement between the calibarated Klinkenberg model and the accurate permeability data is good in the whole range of flow regime, which is different from the conclusion in most of the recent publications, where accurate permeability data at high $Kn$ is not available or not used to calibrate the Klinkenberg model and the model parameter $b$ was empirically determined.

Table 1. Permeability $\kappa$ at $p_0$=0.04 MPa computed using different cell numbers at the $x$ and $y$ directions, respectively, to check the computational error due to low spatial resolution, $Kn$=0.644.

| Cell number | $10^2$ | $20^2$ | $30^2$ | $40^2$ | $50^2$ | $100^2$ |
|---|---|---|---|---|---|---|
| $\kappa$ (×$10^{-15}$ m$^2$) | 9.639 | 10.28 | 10.44 | 10.55 | 10.57 | 10.57 |

Table 2. Permeability $\kappa$ at $p_0$=0.40 MPa computed using different cell numbers at the $x$ and $y$ directions, respectively, to check the computational error due to low spatial resolution, $Kn$=0.0644.

| Cell number | $10^2$ | $20^2$ | $30^2$ | $40^2$ | $50^2$ | $100^2$ | $400^2$ |
|---|---|---|---|---|---|---|---|
| $\kappa$ (×$10^{-15}$ m$^2$) | 1.030 | 1.466 | 1.806 | 2.046 | 2.215 | 2.575 | 2.744 |
| Cell number | 10×100 | 20×50 | 30×40 | 40×30 | 50×20 | 100×10 | |
| $\kappa$ (×$10^{-15}$ m$^2$) | 1.840 | 1.970 | 1.984 | 1.851 | 1.603 | 1.244 | |

Table 3. Permeabilities $\kappa$ computed at different pressure conditions with $\Delta L<\lambda_0$.

| $p_0$ (MPa) | 0.001 | 0.002 | 0.004 | 0.006 | 0.008 | 0.01 | 0.02 | 0.04 | 0.06 |
|---|---|---|---|---|---|---|---|---|---|
| Cell number | $25^2$ | $25^2$ | $25^2$ | $25^2$ | $25^2$ | $25^2$ | $25^2$ | $25^2$ | $25^2$ |
| $Kn$ | 25.78 | 12.89 | 6.444 | 4.296 | 3.222 | 2.578 | 1.289 | 0.644 | 0.430 |
| $\bar{u}$ (m/s) | 0.362 | 0.362 | 0.363 | 0.364 | 0.366 | 0.369 | 0.379 | 0.407 | 0.431 |
| $\kappa$ (×$10^{-15}$ m$^2$) | 370.8 | 185.3 | 93.05 | 62.20 | 46.89 | 37.76 | 19.43 | 10.42 | 7.360 |
| $p_0$ (MPa) | 0.08 | 0.1 | 0.4 | 0.8 | 1.2 | 1.6 | 2.0 | 3.0 | |
| Cell number | $25^2$ | $25^2$ | $100^2$ | $200^2$ | $300^2$ | $400^2$ | $500^2$ | $750^2$ | |
| $Kn$ | 0.322 | 0.258 | 0.064 | 0.032 | 0.021 | 0.016 | 0.013 | 0.0086 | |
| $\bar{u}$ (m/s) | 0.457 | 0.481 | 1.006 | 1.710 | 2.438 | 3.177 | 3.917 | 5.840 | |
| $\kappa$ (×$10^{-15}$ m$^2$) | 5.853 | 4.927 | 2.575 | 2.189 | 2.080 | 2.033 | 2.005 | 1.993 | |

## Appendix

Klinkenberg first studied the gas permeability variation with average pressure $\bar{p}$ (equivalent to $p_0$ here) using the experimental data [2] of rock samples from tight reservoir, where the permeability is at the level of milli-Darcys and thus experimental measurements can be feasible and accurate. The general conclusions are valid for shale rocks having permeability at the level of nano-Darcys. In our work, we conclude that the Klinkenberg correlation model can be used as a fitting formula instead of a physical model and then is always appropriate in the whole flow regime as long as it has been calibrated by determining $\kappa_\infty$ and $b$ using accurate permeability data at the two ends with low and high pressures, respectively. The computed permeability data of a two-dimensional benchmark problem is used to calibrate and verify the Klinkenberg model. Here, we use the experimental data of two real rock samples with large difference of permeability to further validate our conclusion.

Table 4. Experimental permeabilities given in Table 5 of [2] and predicted permeabilities using the Klinkenberg model, where $\kappa_\infty$=2.36 mD and $b$=0.58 atm as given in Table 8 of [2].

| $1/\bar{p}$ (atm$^{-1}$) | 0.052 | 0.083 | 0.154 | 0.308 | 0.397 | 0.594 | 0.839 | 2.167 | 2.859 | 10.19 | 18.17 | 24.96 | 45.56 |
|---|---|---|---|---|---|---|---|---|---|---|---|---|---|
| Exp. $\kappa$ (mD) [2] | 2.420 | 2.480 | 2.640 | 2.900 | 3.040 | 3.430 | 3.760 | 5.640 | 6.690 | 16.65 | 27.96 | 37.13 | 64.84 |
| $\kappa$ (mD) by Eq. (3) | 2.431 | 2.473 | 2.571 | 2.782 | 2.904 | 3.173 | 3.508 | 5.326 | 6.274 | 16.31 | 27.22 | 36.52 | 64.72 |

Table 5. Experimental permeabilities given in Table 6 of [2] and predicted permeabilities using the Klinkenberg model, where $\kappa_\infty$=23.66 mD and $b$=0.116 atm as given in Table 8 of [2].

| $1/\bar{p}$ (atm$^{-1}$) | 0.050 | 0.156 | 0.229 | 0.489 | 1.252 | 2.913 | 6.831 | 11.26 | 16.89 | 42.81 | 61.24 | 84.10 | 96.53 |
|---|---|---|---|---|---|---|---|---|---|---|---|---|---|
| Exp. $\kappa$ (mD) [2] | 23.65 | 24.25 | 25.02 | 26.33 | 28.60 | 34.20 | 46.60 | 60.00 | 75.60 | 148.0 | 197.0 | 259.0 | 290.0 |
| $\kappa$ (mD) by Eq. (3) | 23.80 | 24.09 | 24.29 | 25.00 | 27.09 | 31.65 | 42.41 | 54.56 | 70.01 | 141.1 | 191.7 | 254.5 | 288.6 |

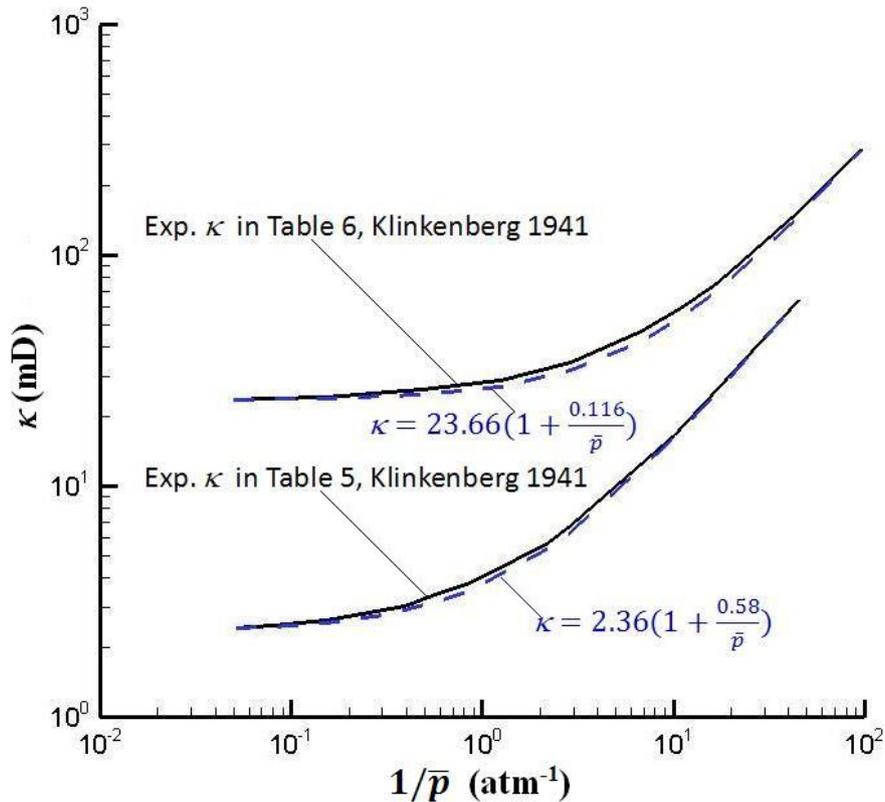

Fig. 4. Comparisons between calibrated Klinkenberg mode and experimental data for two rock samples [2].